\documentclass[9pt, twoside]{article}
\usepackage{japs-ISOSS}

\newcommand{\NN}{\mathbb{N}}

\newcommand{\EE}{\mathbb{E}}

\newcommand{\dd}{\mathrm{d}}
\newcommand{\ee}{\mathrm{e}}

\begin{document}
\thispagestyle{empty} \setcounter{page}{1} \setcounter{section}{0}
\setcounter{equation}{0}\setcounter{theorem}{0}\setcounter{footnote}{0}
\setcounter{table}{0} \setcounter{figure}{0}
\numberwithin{equation}{section}

\vspace{-.20in} \baselineskip .18in

\vspace{.4in}



\title{Fluctuation analysis with cell deaths}

\author{Bernard YCART} 

\address{Univ. Grenoble-Alpes\\ 
Laboratoire Jean Kuntzmann, CNRS UMR 5224\\
51 rue des Math\'ematiques 38041 Grenoble cedex 9, France}

\email{Bernard.Ycart@imag.fr} 




\vspace{.2in}

\centerline{\sc{summary}} \vspace{0.1in} \baselineskip.20in
\centerline{
\begin{minipage}{4.5in} \small{ \noindent
The classical Luria-Delbr\"uck model for fluctuation analysis
 is extended to the case
where cells can either divide or die at the end of their generation
time. This leads to a family of probability distributions generalizing
the Luria-Delbr\"uck family, and depending on three parameters:
the expected number of mutations, the relative fitness of normal
cells compared to mutants, and the
death probability of mutants. The probabilistic treatment is
similar to that of the classical case; simulation and computing
algorithms are provided. The estimation problem is discussed: if the
death probability is known, the two other parameters can be
reliably estimated. If the death probability is unknown, the
model can be identified only for large samples.
}\end{minipage}}

\vspace{.15in} \baselineskip.18in


\centerline{ \begin{minipage}{4.5in} \small{ \noindent {\it Keywords
and phrases:} Bellman-Harris branching
process; cell kinetics; fluctuation analysis; Luria-Delbr\"uck
distribution; mutation model}\end{minipage}}

\vspace{.1in}

\centerline{ \begin{minipage}{4.5in} \small{ \noindent {\it AMS
Classification:} 92D25; 60J28}\end{minipage}}

\section{Introduction}
Since it appeared more than 60 years ago, 
the Luria-Delbr\"uck distribution has been widely used as a model for
the occurrence of mutants in
cell cultures: see chap.~II p.~59 of 
Kendall (1952)
for an early review, and 
Zheng (1999; 2010)
for more recent
ones. It is obtained as a limit when the initial number of cells and
the experiment time are large, and the mutation probability is small.  
One of the underlying hypotheses is that cells only
divide and never die, which is untrue in reality, even though the
probability of death has been estimated to rather
low values 
(Stewart et al. 2005; Fontaine et al. 2008). 
A Markovian model of mutations
including cell deaths was considered by 
Tan (1982), 
who proposed a computing algorithm for the distribution of
mutants. Angerer (2001, section 3)
also discussed the influence of cell
death on the distribution of mutants. To the best of our knowledge no
explicit representation of the distribution of mutants in a general model
including cell deaths, and no quantitive study of the influence of deaths on the
estimation of parameters have appeared so far.
Our objective here
is to extend the 
classical Luria-Delbr\"uck 
model to the case where cells have a certain probability to
die rather than divide, and provide statistical tools for the
estimation of the parameters. 

\noindent
Our hypotheses are the following:
\begin{itemize}
\item at time $0$ a homogeneous population of $n$ normal cells is given;
\item the generation time of any normal cell is a random variable with
  distribution $G$;
\item upon completion of the generation time of a normal cell:
\begin{itemize}
\item with probability $p$ one normal and one mutant cell are produced;
\item with probability $q$ the cell dies out;
\item with probability $1-p-q$ two normal cells are produced,
\end{itemize}
\item the generation time of any mutant cell is exponentially
  distributed with parameter $\mu^*$;
\item upon completion of the generation time of a mutant cell:
\begin{itemize}
\item with probability $\delta$ the cell dies out;
\item with probability $1-\delta$ two mutant cells are produced,
\end{itemize}
\item all random variables and events (division times, mutations, and deaths) 
are mutually independent.
\end{itemize}
Consider an initial (large) number $n$ of normal cells. Assume
that the mutation probability $p$ is small, that the time $t$
at which mutants are counted is large, and that the asymptotics are
such that the expected number of mutations $\alpha$ before
time $t$ is non null and finite (precise hypotheses and statements 
will be given in section 
\ref{asymptotics}). Denote by $\nu$ and $\mu$ the exponential growth rates of
normal and mutant populations respectively, and by $\rho=\nu/\mu$ the
\emph{relative fitness}. 
It will be shown that the total number of mutants at time $t$ approximately
follows an integer valued distribution, whose probability 
generating function (PGF) is given by:
\begin{equation}
\label{ldd}
g_{\alpha,\rho,\delta}(z) = \exp(\alpha(h_{\rho, \delta}(z)-1))\;,
\end{equation}
where:
\begin{equation}
\label{yd}
h_{\rho,\delta}(z)=\int_0^1 \frac{\delta(1-z)+
v\left((1-\delta)z-\delta\right)}
{(1-\delta)(1-z)+v\left((1-\delta)z-\delta\right)}\,\rho v^{\rho-1}\,\dd v\;. 
\end{equation}
The parameters are:
\begin{enumerate}
\item $\alpha$: the expected number of mutations
\item $\rho$: the relative fitness of normal cells compared to
  mutants.
\item $\delta$: the death probability of a mutant cell.
\end{enumerate}
For $\delta=0$, the Luria-Delbr\"uck distribution with
parameters $\alpha$ and $\rho$, or LD$(\alpha,\rho)$, is obtained as a
particular case. We 
shall name ``Luria-Delbr\"uck with deaths'', and denote by
LDD$(\alpha,\rho,\delta)$, the distribution on integers
with PGF $g_{\alpha,\rho,\delta}$. This paper proposes
a statistical study of the LDD$(\alpha,\rho,\delta)$ including:
\begin{itemize}
\item fast simulation algorithm,
\item computation of probabilities,
\item asymptotic tail probabilities,
\item point estimation of parameters,
\item confidence intervals.
\end{itemize}
We have developed in R 
(R core development team 2008)
a set of functions  that perform the usual
operations on the LDD distributions (simulation, distribution function
and quantile computation), output estimates and confidence intervals. 
These functions have been made available 
online: \texttt{http://www.ljk.imag.fr/membres/Bernard.Ycart/LD/}.
\\
The paper is organized as follows. In section \ref{asymptotics}, 
the theoretical justification of the model is presented. 
It is based on standard
results from branching process theory. 
A simple probabilistic interpretation will be
given. Section \ref{computation} describes the simulation and 
computation algorithms of the LDD$(\alpha,\rho,\delta)$: they are
quite similar to those known for the LD$(\alpha,\rho)$
(Zheng 2005).
The estimation problem is addressed in section \ref{estimation}. The
proposed method is based on generating function estimates, extending
those defined for the LD$(\alpha,\rho)$ in 
(Hamon and Ycart 2012). 
Experimental results, both on simulated and real data are reported in
section \ref{experiments}.
\section{Asymptotics for numbers of mutants}
\label{asymptotics}
The results exposed in this section are applications of the
general theory of supercritical age-dependent continuous time
branching processes (or Bellman-Harris processes): see 
Chap. VI of Harris (1963) and Chap. IV of Athreya and Ney (1972)
as general references. They are similar to those detailed in section 2
of Hamon and Ycart (2012),
and we shall mainly develop differences
with the classical model, that arise when taking
cell deaths into account.

Firstly, consider the number of normal cells as a function of
time. Recall that the generation times are assumed to be independent
and identically distributed (i.i.d.) random variables, with common
distribution $G$. Upon completion of a generation time, the number of
(normal) offspring is:
\begin{itemize}
\item[0] with probability $q$ (death),
\item[1] with probability $p$ (mutation),
\item[2] with probability $1-p-q$ (division).
\end{itemize}  
Therefore the PGF of the offspring distribution is:
$$
q+pz+(1-p-q)z^2\;,
$$
and its expectation (mean offspring number) is $m=p+2(1-p-q)$. We
shall assume that the mean offspring number is larger than 1, so the
corresponding clones may survive with positive probability 
(supercritical case). If one
normal cell is initially present, then either the population dies out
or it grows exponentially. The probability that it dies out is the
smallest positive root of the equation $z=q+pz+(1-p-q)z^2$, which we
shall denote by $\varepsilon$:
$$
\varepsilon = \frac{q}{1-p-q}\;. 
$$
With probability $1-\varepsilon$ the clone does not die out, in which
case it grows exponentially. The exponential growth rate (or
\emph{Malthusian parameter}) $\nu$ 
is  defined as the unique root of the equation:
\begin{equation}
\label{malthus}
m\int_0^{+\infty} \ee^{-\nu s}\,\dd G(s)=1\;.
\end{equation}
Theorem 17.1 p.~142 of 
Harris (1963)
gives a precise
meaning to the expression ``exponential growth''.
It states that:
$$
\lim_{t\to+\infty} \EE[N_t\,|\,N_0=1,\,N_t>0]\,\ee^{-\nu t} = 
C\;,
$$
where $N_t$ denotes the number of normal cells at time $t$.
The limit is the proportionality coefficient of exponential
growth. It is given by:
\begin{equation}
\label{propor}
C= \left(\nu\frac{m^2}{m-1}\int_0^{+\infty} s\ee^{-\nu s}\,\dd G(s)\right)^{-1}\;.
\end{equation}
Assume $n$ normal cells are present at time $t=0$. 
Let $(t_n)$ be a sequence of instants, tending to infinity as $n$
tends to infinity. At large time $t_n$,
a proportion $\varepsilon$ of the clones stemming from the $n$ initial
cells will have died out. A proportion $1-\varepsilon$ grow
exponentially with rate $\nu$. So the final number of normal cells
will be asymptotically equivalent to $n(1-\varepsilon)C\ee^{\nu t_n}$.

Consider now mutations. Let $(p_n)$ be a sequence
of mutation probabilities, tending to $0$ as $n$ tends to infinity. Since
$(p_n)$ tends to zero, mutations have an asymptotically null effect on
the growth rate of the population. Indeed the mean offspring number tends to
$2(1-q)$, the growth rate tends to the unique solution $\nu$ of
the equation:
$$
2(1-q)\int_0^{+\infty} \ee^{-\nu s}\,\dd G(s)=1\;,
$$
and the proportionality constant tends to:
$$
C= \left(\nu\frac{4(1-q)^2}{1-2q}
\int_0^{+\infty} s\ee^{-\nu s}\,\dd G(s)\right)^{-1}\;.
$$
Moreover, since the number of divisions occurring in dying
clones remains bounded, they can be neglected, and it can be considered that
mutants observed at time $t_n$ only come from divisions in surviving
mutant clones. Their number is asymptotically equal to the final
number of normal cells, i.e. 
$n(1-\varepsilon)C\ee^{\nu t_n}$. Assume now that:
\begin{equation}
\label{hypalpha}
\lim_{n\to+\infty} p_nn(1-\varepsilon)C\ee^{\nu t_n}=\alpha\;,
\end{equation}
where $\alpha$ is some fixed positive real. The expected number of
mutations tends to $\alpha$, and since mutations are supposed to occur
randomly, their number asymptotically follows the Poisson distribution
with parameter $\alpha$, by the law of small numbers. Notice that the
interpretation of $\alpha$ as the product of the mutation probability
by the final number of cells holds whether cell deaths are considered or not; 
thus estimating $\alpha$ permits to estimate the mutation
probability $p$, dividing the estimate of $\alpha$ by the final number
of cells, exactly as in classical fluctuation analysis. 

Consider now the durations between random split times of non dying
clones, and the final time $t_n$: we call them \emph{split lags}. 
Theorem 2.1 p.~669 of
Kuczek (1982)
states the almost sure convergence of the
empirical distribution of split lags,
to the distribution function of the exponential with parameter $\nu$.
From section 3 of that same article, it follows that the developing
times of a fixed number $k$ 
of mutant clones converge in distribution to the product of $k$
independent copies of the exponential distribution with
parameter $\nu$.

Let us now turn to mutant clones, i.e. populations of mutant cells
stemming from a single initial mutant cell. Recall that the generation times
of mutants are assumed to be exponentially distributed with rate
$\mu^*$. The number of mutants at time $s$ is a linear growth
birth-and-death process. The 
rates are proportional to the number $k$ of cells in the population,
the death rate being $\mu^*\delta k$ and the birth rate being
$\mu^*(1-\delta) k$. We shall assume also that mutant clones may
survive with positive probability, which occurs only if $\delta<1/2$. 
The exponential growth rate of mutant clones
is the difference between birth and death individual rates:
$$
\mu = \mu^*(1-\delta)-\mu^*\delta=\mu^*(1-2\delta)\;.
$$
The distribution at time $s$ of the number of mutant cells, stemming
from a single mutant cell at time $0$ is explicitly known, and
characterized by the following PGF:
(Athreya and Ney 1972, p.~109).
\begin{equation}
\label{gfmutantclone}
F(z,s) = \frac{\delta(1-z)+\ee^{-\mu t}((1-\delta)z-\delta)}
{(1-\delta)(1-z)+\ee^{-\mu t}((1-\delta)z-\delta)}\;.
\end{equation}
Let us summarize the 3 main arguments:
\begin{itemize}
\item[A1:]
the number of mutations converges in distribution to 
the Poisson distribution with parameter $\alpha$;
\item[A2:] 
the joint distribution of the developing times of a fixed number $k$
of mutant clones converges in distribution to the product of $k$
independent copies of the exponential distribution with parameter $\nu$;
\item[A3:]
the size at time $s$ of a mutant clone has 
distribution with PGF $F(z,s)$.
\end{itemize}
From A2, the size of any mutant
clone is an exponential mixture, the PGF of which can
be expressed using (\ref{gfmutantclone}) as:
$$
\int_0^{+\infty}
\frac{\delta(1-z)+\ee^{-\mu t}((1-\delta)z-\delta)}
{(1-\delta)(1-z)+\ee^{-\mu t}((1-\delta)z-\delta)}
\,\nu\ee^{-\nu s}\,\dd s\;.
$$
Changing $\ee^{-\mu s}$ into $v$ yields the expression (\ref{yd}) of 
$h_{\rho,\delta}$. The distribution with PGF
$h_{\rho,\delta}$ can be seen as a two-parameter extension of the Yule
distribution with parameter $\rho$: it will be denoted by
YD$(\rho,\delta)$ (for ``Yule with deaths'').
From A1, the total number of mutants is the sum of a random
number of sizes of independent random clones, each with
YD$(\rho,\delta)$ distribution: the resulting distribution is a
compound Poisson with parameter $\alpha$ and base YD$(\rho,\delta)$,
hence the expression  (\ref{ldd}) of the PGF
$g_{\alpha,\rho,\delta}$ of the LDD$(\alpha,\rho,\delta)$.
\section{Probability calculations}
\label{computation}
Computation and simulation algo\-rithms for the YD$(\rho,\delta)$ and the 
LDD$(\alpha,\rho,\delta)$ distributions are described in this section. 
The probabilities of the YD$(\rho,\delta)$ and LDD$(\alpha,\rho,\delta)$
will be denoted by $(p_k)_{k\in\NN}$ and $(q_k)_{k\in\NN}$  respectively.
$$
h_{\rho,\delta}(z) = \sum_{k=0}^{+\infty} p_k\,z^k
\quad\mbox{and}\quad
g_{\alpha,\rho,\delta}(z) = \sum_{k=0}^{+\infty} q_k\,z^k\;.
$$

We begin with a probabilistic interpretation of the distribution 
at time $s$ of mutant clones, the PGF $F(z,s)$ 
of which is given by (\ref{gfmutantclone}). Let us rewrite $F(z,s)$ as:
$$
F(z,s)=\frac{\delta(1-\ee^{-\mu s})-z(\delta-\ee^{-\mu s}(1-\delta))}
{(1-\delta-\delta\ee^{-\mu s})-z((1-\delta)(1-\ee^{-\mu s})}\;.
$$
An easy series expansion yields the corresponding probabilities:
$$
F(z,s) = \sum_{k=0}^{+\infty} p_k(s)\,z^k\;,
$$
with
$$
p_0(s)=\frac{\delta(1-\ee^{-\mu s})}{1-\delta-\delta\ee^{-\mu s}}\;,
$$
and for $k\geqslant 1$,
\begin{equation}
\label{pks}
p_k(s) = (1-p_0(s))\pi(s)(1-\pi(s))^{k-1}
\mbox{, with } 
\pi(s) = \frac{(1-2\delta)\ee^{-\mu s}}{1-\delta-\delta \ee^{-\mu s}}\;.
\end{equation}
In other words, a random variable with PGF $F(z,s)$ is
a random mixture: either $0$ with probability $p_0(s)$ or (with
probability $1-p_0(s)$), a geometric
random variable with parameter $\pi(s)$.  
The YD$(\rho,\delta)$ is an exponential mixture of these
distributions. Using again the change of variable $\ee^{-\mu s}\mapsto
v$,
\begin{equation}
\label{p0}
p_0 = \int_0^{1} \frac{\delta(1-v)}{1-\delta-\delta v}\,
\rho v^{\rho-1}\,\dd v\;,
\end{equation}
and for $k\geqslant 1$:
\begin{equation}
\label{pk}
p_k = (1-\delta)^{k-1}(1-2\delta)^2
\int_0^{1} \frac{(1-v)^{k-1}}{(1-\delta-\delta v)^{k+1}}\,
\rho v^{\rho}\,\dd v\;.
\end{equation}
The integral in (\ref{pk}) can be computed numerically up to rather
large values of $k$. An equivalent for larger $k$'s can be calculated as
follows. Rewrite (\ref{pk}) as:
\begin{eqnarray*}
p_k &=& \displaystyle{
\left(\frac{1-2\delta}{1-\delta}\right)^2
\int_0^{1} \frac{(1-v)^{k-1}}{(1-v\frac{\delta}{1-\delta})^{k+1}}
\,\rho v^{\rho}\,\dd v
}\\[2ex]
&=&\displaystyle{k^{-\rho-1}
\left(\frac{1-2\delta}{1-\delta}\right)^2
\int_0^{k} \frac{(1-\frac{u}{k})^{k-1}}
{(1-\frac{u}{k}\frac{\delta}{1-\delta})^{k+1}}
\,\rho u^{\rho}\,\dd u
}
\end{eqnarray*}
The following equivalent is obtained.
\begin{equation}
\label{pinf}
p_k \mathop{\sim}_{k\to \infty}
k^{-\rho-1}\left(\frac{1-2\delta}{1-\delta}\right)^{1-\rho}\rho\,\Gamma(\rho+1)\;.
\end{equation}
A well known algorithm expresses the $q_k$'s as a function
of the $p_k$'s: see 
Embrecht and Hawkes (1982) or Pakes (1993).
\begin{equation}
\label{algo}
q_0=\ee^{-\alpha(1-p_0)},
\quad\mbox{and for $k\geqslant 1$,}\quad
q_k = \frac{\alpha}{k} \sum_{i=1}^k ip_i q_{k-i}\;.
\end{equation}
The proof of (\ref{algo}) is easy:
\begin{eqnarray*}
\displaystyle{\frac{\dd g_{\alpha,\rho,\delta}}{\dd z}} &=&
\displaystyle{ 
\alpha\frac{\dd h_{\rho,\delta}}{\dd z}g_{\alpha,\rho,\delta}}\\[2ex]
&=&\displaystyle{\alpha\left(\sum_{i=1}^{+\infty}ip_iz^{i-1}\right)
\left(\sum_{k=0}^{+\infty}q_kz^{k}\right)}\\[2ex]
&=&\displaystyle{\sum_{k=1}^{+\infty}kq_kz^{k-1}}\;.
\end{eqnarray*}
The equivalent of $q_k$ is deduced from 
subexponential theory: 
(Embrecht and Kawkes 1982, Theorem 1).
\begin{equation}
\label{qinf}
q_k \mathop{\sim}_{k\to \infty} \alpha p_k
\mathop{\sim}_{k\to \infty} \alpha
k^{-\rho-1}\left(\frac{1-2\delta}{1-\delta}\right)^2\rho\,\Gamma(\rho+1)\;.
\end{equation}
For any $\delta$, a heavy tail distribution with tail
exponent $\rho$ is obtained. 

Another consequence of the probabilistic interpretation 
 is a simulation algorithm
for the YD$(\rho,\delta)$ and the LDD$(\alpha,\rho,\delta)$. 

\noindent
For the YD$(\rho,\delta)$:
\begin{itemize}
\item simulate a random time $s$, according to the exponential
  distribution with parameter $\rho$; 
\item compute $p_0(s)$ and $\pi(s)$;
\item make a random choice:
\begin{itemize}
\item with probability $p_0(s)$, output $0$,
\item with probability $1-p_0(s)$, output a geometric random number
  with parameter $\pi(s)$. 
\end{itemize}
\end{itemize}
For the LDD$(\alpha,\rho,\delta)$:
\begin{itemize}
\item simulate a random integer $n$ according to the Poisson
  distribution with parameter $\alpha$; 
\item simulate a sample of size $n$ of the YD$(\rho,\delta)$,
\item output the sum of that sample.
\end{itemize}
These two algorithms have been encoded in R, and the simulation functions
are included in the script available online.
\section{Parameter estimation}
\label{estimation}
This section addresses the problem of parameter estimation. 
The main difficulty comes from the fact that two LDD distributions
may be quite close for rather different sets of parameters; this
makes the model hardly identifiable in practice. In order to evaluate 
the actual identifiability, we proceed as follows. Let $\alpha_0$ and
$\rho_0$ be two given positive values, and let $(q_0,q_1)$
be the first two probabilities of the LDD$(\alpha_0,\rho_0,0)$.
For any value $\delta<0.5$, there exists a couple 
$(\alpha_\delta,\rho_\delta)$ such that the first two probabilities of
the LDD$(\alpha_\delta,\rho_\delta,\delta)$ coincide with $(q_0,q_1)$.
It turns out that the whole distribution 
LDD$(\alpha_\delta,\rho_\delta,\delta)$ is very close to
the LDD$(\alpha_0,\rho_0,0)$. 
Let $\mathrm{dist}(\delta)$ be the maximal 
distance between the two cumulative distribution functions. 
Table \ref{tab:distancesq0q1} gives the values of $\alpha_\delta$,
$\rho_\delta$,  
and $\mathrm{dist}(\delta)$ for $\alpha_0=\rho_0=1$
and $\delta$ from $0$ to $0.3$.
\begin{table}[!ht]
$$
\begin{array}{|c|ccccccccccc|}
\hline
\delta& 0& 0.03& 0.06& 0.09& 0.12& 0.15& 0.18& 0.21& 0.24&  0.27&  0.30\\
\alpha_\delta&1& 1.02& 1.03& 1.05& 1.08& 1.10& 1.13& 1.16& 1.20&  1.25&  1.30\\
\rho_\delta&  1& 1.01& 1.02& 1.04& 1.05& 1.07& 1.09& 1.12& 1.15&  1.19&  1.24\\
10^3\,\mathrm{dist}(\delta)&0& 0.64& 1.35& 2.12& 2.98& 3.94&
5.02& 6.24& 7.63&  9.24& 11.11\\
\hline
\end{array}
$$
\caption{Parameters of LDD distributions that coincide with
LDD$(1,1,0)$ on 0 and 1,
  with maximal distance between cumulative distribution functions,
  multiplied by $10^3$.}
\label{tab:distancesq0q1}
\end{table}

Of course, $\mathrm{dist}(\delta)$ depends on $\alpha_0$, $\rho_0$ and
$\delta$:  
it increases with $\alpha_0$ and $\delta$, it decreases with $\rho_0$; 
but its typical order of magnitude is $10^{-3}$. As a consequence,
there is no hope to distinguish
between LDD distributions on a sample of a few hundred data, which is 
the usual size in fluctuation analysis experiments. However,
the observed number of mutants increases with $\alpha$ (the expected
number of mutations), and so does
the identifiability of the model. 
Here are the values of $\mathrm{dist}(\delta)$ for $\delta=0.1$, 
$\rho_0=1$, and $\alpha_0$ from $10$ to $50$.
$$
\begin{array}{|c|ccccc|}
\hline
\alpha_0&10& 20& 30& 40& 50\\
10^3\, \mathrm{dist}(0.1)&
14.30& 21.91& 27.52& 32.06& 35.91
\\\hline
\end{array}
$$
The fact that more precise estimates should be obtained for large
values of $\alpha$ rules out in our view the maximum likelihood
method, as already argued in 
Hamon and Ycart (2012). It is
the main argument supporting empirical probability generating function
(GF) estimators. 
Since the pioneering work of Parzen (1962), 
estimators
based on empirical exponential transforms (characteristic function, moment
generating function, probability generating function) have been widely
used, in particular for heavy tail distributions: see Yao and Morgan
(1999), 
Yu (2004)
for general reviews,
Dowling and Nakamura (1997) 
for GF
estimators. Remillard and Theodorescu (2000) 
treat a case similar to ours. The
estimators defined below extend those introduced for the
LD$(\alpha,\rho)$ in 
Hamon and Ycart (2012).

Recall the PGF of the LDD$(\alpha,\rho,\delta)$:
$$
g_{\alpha,\rho,\delta}(z)=\exp(-\alpha (1-h_{\rho,\delta}(z)))\;.
$$
For $0\leqslant \delta<0.5$ and $0\leqslant z<1$, we shall denote by
$\delta_*$ and $z_*$ the following quantities.
$$
\delta_* = \frac{\delta}{1-\delta}
\quad\mbox{and}\quad
z_* = \frac{z-\delta_*}{1-z}\;.
$$
The PGF  $h_{\rho,\delta}$ and its derivatives
with respect of $\rho$ and $\delta$, denoted by
$h^{(\rho)}_{\rho,\delta}(z)$ and $h^{(\delta)}_{\rho,\delta}(z)$
respectively, 
are repeatedly needed in algorithmic
procedures, so numerically stable expressions must be derived. Here
are the expressions that  have
been implemented in our R functions.
\begin{equation}
\label{hstable}
h_{\rho,\delta}(z) = \delta_*+
z_*(1-\delta_*)\int_0^1\frac{\rho v^\rho}{1+z_*v}\,\dd v\;.
\end{equation}
\begin{equation}
\label{hrhostable}
h^{(\rho)}_{\rho,\delta}(z)
=\frac{\partial h_{\rho,\delta}(z)}{\partial\rho}
= 
z_*(1-\delta_*)\int_0^1\frac{v^\rho}{1+z_*v}(1+\rho\log(v))\,\dd v\;.
\end{equation}
\begin{equation}
\label{hdeltastable}
\begin{array}{lcl}
\displaystyle{h^{(\delta)}_{\rho,\delta}(z)
=\frac{\partial h_{\rho,\delta}(z)}{\partial\delta}}
&=& 
\displaystyle{\left(\frac{1}{1-\delta}\right)^2\left(1-
\left(z_*+\frac{1-\delta_*}{1-z}\right)
\int_0^1\frac{\rho v^\rho}{1+z_*v}\,\dd v\right.}\\[2ex]
&&\displaystyle{\hspace*{2.5cm}\left.+\frac{z_*(1-\delta_*)}{1-z}
\int_0^1\frac{\rho v^{\rho+1}}{(1+z_*v)^2}\,\dd v\right)\;.}
\end{array}
\end{equation}
We use a method of moments, such as stated by
R\'emillard and Theodorescu (2000)
in a similar context.
Let $0<z_1<z_2<z_3<1$ be three different values, considered as
fixed. Let $g_1,g_2,g_3$ be their respective images by
$g_{\alpha,\rho,\delta}$. Denote by $G=G(\alpha,\rho,\delta)$ the 
3-dimensional vector $(g_1,g_2,g_3)$. 
The mapping 
$(\alpha,\rho,\delta)\longmapsto G$ is locally one-to-one, and its
inverse can be used to derive an estimate of $(\alpha,\rho,\delta)$ from the
natural estimate of $G$. 
Let $(X_n)_{n \geqslant 1}$ be a sequence of independent identically 
distributed random variables, each with PGF 
$g_{\alpha,\rho,\delta}$. Define the
\emph{empirical probability generating function} (EPGF) $\hat{g}_n(z)$ as:
$$
\hat{g}_n(z) = \frac{1}{n} \sum_{i=1}^n z^{X_i}\;.
$$
For $i,j=1,2,3$, the expectations and covariances of $\hat{g}_n(z_i)$
and $\hat{g}_n(z_j)$ are easily expressed:
$$
\EE[\hat{g}_n(z_i)]=g_{\alpha,\rho,\delta}(z_i)\;,
$$
and
$$
\mbox{cov}[\hat{g}_n(z_i),\hat{g}_n(z_j)]=c(z_i,z_j)=
g_{\alpha,\rho,\delta}(z_iz_j)
-
g_{\alpha,\rho,\delta}(z_i)
g_{\alpha,\rho,\delta}(z_j)\;.
$$ 
Consider the 3-dimensional vector
$$
\hat{G}=(\hat{g}_n(z_1),
\hat{g}_n(z_2),
\hat{g}_n(z_3))\;.
$$
Its coordinates are empirical means of independent, identically distributed,
bounded random variables: hence it is a strongly consistent estimator
of $G$. By the Central Limit 
Theorem,  $\sqrt{n}(\hat{G}-G)$ converges in distribution to the
trivariate centered normal 
distribution, with covariance matrix
$C=(c(z_i,z_j))_{i,j=1,2,3}$.
R\'emillard and Theodorescu (2000, Proposition 3.1)
give a stronger result,
stating the functional
convergence of $\hat{g}_n(z)$ to a Gaussian process. 

The Jacobian matrix of $G$ as a function of 
$(\alpha,\rho,\delta)$ is the following.
\begin{eqnarray*}
J &=& 
\left(\begin{array}{ccc}
\frac{\partial g_{\alpha,\rho,\delta}(z_1)}{\partial\alpha}&
\frac{\partial g_{\alpha,\rho,\delta}(z_1)}{\partial\rho}&
\frac{\partial g_{\alpha,\rho,\delta}(z_1)}{\partial\delta}\\
\frac{\partial g_{\alpha,\rho,\delta}(z_2)}{\partial\alpha}&
\frac{\partial g_{\alpha,\rho,\delta}(z_2)}{\partial\rho}&
\frac{\partial g_{\alpha,\rho,\delta}(z_2)}{\partial\delta}\\
\frac{\partial g_{\alpha,\rho,\delta}(z_3)}{\partial\alpha}&
\frac{\partial g_{\alpha,\rho,\delta}(z_3)}{\partial\rho}&
\frac{\partial g_{\alpha,\rho,\delta}(z_3)}{\partial\delta}
\end{array}\right)\\[3ex]
&=&
\left(\begin{array}{ccc}
g_1(h_{\rho,\delta}(z_1)-1)&
g_1\alpha h^{(\rho)}_{\rho,\delta}(z_1)&
g_1\alpha h^{(\delta)}_{\rho,\delta}(z_1)\\
g_2(h_{\rho,\delta}(z_2)-1)&
g_2\alpha h^{(\rho)}_{\rho,\delta}(z_2)&
g_2\alpha h^{(\delta)}_{\rho,\delta}(z_2)\\
g_3(h_{\rho,\delta}(z_3)-1)&
g_3\alpha h^{(\rho)}_{\rho,\delta}(z_3)&
g_3\alpha h^{(\delta)}_{\rho,\delta}(z_3)\\
\end{array}\right)
\end{eqnarray*}
 Admitting that $J$ is non singular,
denote by $\phi$ the inverse of the mapping
$(\alpha,\rho,\delta)\longmapsto G$. Then $\phi(\hat{G})$ is a
consistent estimator of $(\alpha,\rho,\delta)$. By Slutsky's
theorem, such as formulated by 
R\'emillard and Theodorescu (2000, Theorem 3.4)
$\sqrt{n}(\phi(\hat{G})-(\alpha,\rho,\delta))$
converges in distribution to the trivariate centered normal
distribution with covariance matrix $(J^{-1})^t\,C\,J^{-1}$. 
From there,
confidence intervals and p-values of hypothesis testing can be
obtained by standard procedures 
(see e.g. Anderson (2003)).
As was explained in 
Hamon and Ycart (2012),
the main advantage of GF
estimators is to allow a rescaling of the sample, which makes the
method applicable to large values of $\alpha$. The idea is to replace 
$z$ by $z^{1/b}$ in the definition of $\hat{g}_n(z)$:
$$
\hat{g}_{n}(z^{1/b})=
\frac{1}{n}\sum_{i=1}^n (z^{1/b})^{X_i}= 
\frac{1}{n}\sum_{i=1}^n z^{X_i/b}\;.
$$ 
The estimator $\phi(\hat{G})$ is theoretically consistent. However,
for intrinsic reasons that were explained at the beginning of this
section, it is numerically unstable, and can
only be applied to very large samples, beyond the size of
those usually collected in
fluctuation analysis experiments. Thus we have been led to propose
other estimators, that will now be described.

We first assume that $\delta$ is known. 
Observe that for $z=\delta_*$, $z_*=0$ and 
$h_{\rho,\delta}(\delta_*)=\delta_*$: $h_{\rho,\delta}$ 
has a fixed point at $\delta_*$, independently of $\rho$.
Therefore 
$\hat{g}_n(\delta_*)$ converges to $g_{\alpha,\rho,\delta}(\delta_*)=
\exp(\alpha(\delta_*-1))$. Hence 
$\log(\hat{g}(\delta_*))/(\delta_*-1)$ is a consistent
estimator of $\alpha$, that we shall call the \emph{fixed point
  estimator}. It does not depend on $\rho$.
For $\delta=0$, the fixed point estimator is 
$-\log(\hat{g}(0))$, and $\hat{g}(0)$ is the
proportion of zeros in the sample. Thus the fixed point estimator
extends the so called $p_0$-method,
initially proposed by 
Luria and Delbruck (1943)
 (see 
Foster (2006)
for
a review on estimation methods for the LD$(\alpha,\rho)$). The
$p_0$-method, even though it gives acceptable results for low values
of $\alpha$, cannot
be applied for large $\alpha$'s: the same can be said of the fixed
point estimator. 

The best results were obtained for the 
GF estimators that were developed in 
Hamon and Ycart (2012)
 for the
LD$(\alpha,\rho)$ case. We briefly recall their definition below.

Consider the following ratio.
$$
f_{z_1,z_2}(\rho)=\frac{h_{\rho,\delta}(z_1)-1}{h_{\rho,\delta}(z_2)-1}\;.
$$
The function that maps $\rho$ onto 
$y=f_{z_1,z_2}(\rho)$ is continuous and strictly monotone, hence
one-to-one. Therefore the inverse, that maps $y$ onto
$\rho=f^{-1}_{z_1,z_2}(y)$, is well defined.
For $0<z_1<z_2<1$, let $\hat{y}_n(z_1,z_2)$ denote the following log-ratio.
$$
\hat{y}_n(z_1,z_2) = 
\frac{\log(\hat{g}_n(z_1))}{\log(\hat{g}_n(z_2))}\;.
$$
An estimator of $\rho$ is obtained by:
$$
\hat{\rho}_n(\delta) = f^{-1}_{z_1,z_2}(\hat{y}_n)\;,
$$
then an estimator of $\alpha$ by:
$$
\hat{\alpha}_n(\delta) = \frac{\log(\hat{g}_n(z_3))}
{h_{\hat{\rho}_n(z_1,z_2)}(z_3)-1}\;.
$$
The asymptotic covariance matrix of $(\hat{\alpha}_n,\hat{\rho}_n)$
given in Proposition 4.1 of 
Hamon and Ycart (2012)
 is still valid here,
replacing $h_\rho$ and its derivative in $\rho$ by $h_{\rho,\delta}$
and $h^{(\rho)}_{\rho,\delta}$.
If we assume now that $\rho$ is known and $\delta$ unknown,
the estimators described above are easily adapted, by 
exchanging the roles of $\rho$ and
$\delta$, and replacing $h^{(\rho)}_{\rho,\delta}$ by
$h^{(\delta)}_{\rho,\delta}$. New GF estimators $\hat{\alpha}_n(\rho)$
and $\hat{\delta}_n(\rho)$ are obtained.

In practice, neither $\delta$ nor $\rho$ can be supposed to be
known. For a given value of $\delta$, consider the estimators
$\hat{\alpha}(\delta)$ and $\hat{\rho}(\delta)$ described above. 
The distributions
LDD$(\hat{\alpha}(\delta),\hat{\rho}(\delta),\delta)$ from different
values of $\delta$ are not far from each other. 
To distinguish between them, we propose to use as an
estimator of $\delta$ the value $\hat{\delta}$ 
that minimizes the distance between
the theoretical PGF and the EPGF 
of the sample (up to possible rescaling). We shall denote by
$\hat{B}=(\hat{\alpha}(\hat{\delta}),\hat{\rho}(\hat{\delta}),\hat{\delta})$
this new estimator. Unlike $\phi(\hat{G})$ 
(which is numerically unstable for small samples),
$\hat{B}$ can be calculated on samples of any size. It will be shown
in the next section that when both can
be calculated, $\hat{B}$ has a better mean squared error (MSE) 
than $\phi(\hat{G})$.

A well known drawback of empirical generating function methods is
to depend on tuning values (the points where the empirical
transform is evaluated), whose optimal setting depends of the parameters
to be estimated, and is therefore unknown. The question of tuning has
been discussed for instance in Brockwell and Brown (1981, section 3)  
for continuous distributions, Dowling and Nakamura (1997)
for discrete distributions. In our case, the three values $z_1,
z_2, z_3$ obviously depend at least 
on $\alpha$: the larger $\alpha$, the larger
$z_1, z_2, z_3$ should be. This is why the scaling parameter $b$ was
introduced, replacing $z_i$ by 
$z_i^{1/b}$. We decided to choose for $b$ the $q$\textsuperscript{th}
quantile of the sample (or $b=1$ in case that quantile is null). 
Thus our estimators depend on a set of four tuning
values: $z_1,z_2,z_3,q$. Given a set $\alpha,\rho,\delta$ 
to be estimated, different target functions can be
chosen for the optimization of the tuning set. 
Following Dowling and Nakamura, 
the determinant of the asymptotic covariance matrix was minimized. Numerical
evidence showed a large variability of the optimal tuning parameters
as a function of $\alpha,\rho,\delta$: results comparable to 
Dowling and Nakamura (1997, section 3) were obtained. We compared them with
a simulation study: $1000$ samples of size $100$ of the
LDD$(\alpha,\rho,\delta)$ were simulated, estimates were calculated
for various tuning sets, and the MSE was
minimized. In many cases, the results of the two minimizations were quite 
different. Our goal was 
to propose a tuning set valid for the widest possible range of
parameters. Based on the simulation study, we settled on the same
default set as in Hamon and Ycart (2012): $z_1=0.1$,
$z_2=0.9$, $z_3=0.8$, $q=0.1$. On simulation experiments of $1000$
samples of size $100$, the coverage probability of $95\%$ confidence
intervals remained acceptable for values of $\alpha$ between $0$ and
$10$, values of $\rho$ between $0.5$ and $4$, values of $\delta$ between
$0$ and $0.3$. We believe that these ranges cover most cases of
practical interest. Moreover as was shown in Hamon and Ycart (2012),
that tuning set yields close to optimal efficiency when compared to
the maximum likelihood estimator in the cases where it can be
computed ($\alpha$ small, $\delta=0$).
\section{Experimental results}
\label{experiments}
Using extensively the simulation procedure described
in section \ref{computation}, we have conducted different simulation
experiments in order to assess the qualities of the estimators
proposed in section \ref{estimation}. We have also used the most comprehensive
data set available so far (1102 values), that of
Boe et al. (1994).
Our main conclusions are
reported in this section.

The reason why the GF estimator $\phi(\hat{G})$ cannot
be calculated for small samples was explained in the previous
section. The question arises to compare it, on large
samples and for large values of $\alpha$, to the GF estimator
$\hat{B}$
obtained by estimating first $\alpha$ and $\rho$ on different values
of $\delta$, and then selecting the set of parameters that minimizes the
distance between PGF's. The second one consistently
gives better results. Here are for instance the MSEs 
on the estimation of the three parameters,  on
1000 simulated samples of size $10^4$ of the LDD$(10,1,0.1)$. 
$$
\begin{array}{|c|ccc|}
\hline
&\alpha&\rho&\delta\\\hline
\phi(\hat{G})&  0.551& 0.019& 0.111\\
\hat{B}& 
0.481& 0.016& 0.079\\\hline
\end{array}
$$ 
Notice that both estimators perform quite poorly on the estimation of
$\delta$: the MSE is comparable to the true value.
As explained in the previous section, this must be blamed on the intrinsic
lack of identifiability of the model, rather than the estimators.

We then tried to evaluate the quality of different estimators of
$\alpha$, which is the parameter of interest in fluctuation
analysis. Four estimators were tried on 1000 samples of size 1000 of
the LDD$(\alpha,\rho,\delta)$, 
computing for each estimator the MSE on $\alpha$.
\begin{itemize}
\item GFd: the estimate $\hat{\alpha}(\delta)$ obtained using the true
  value of $\delta$;
\item GFr: the estimate $\hat{\alpha}(\rho)$ obtained using the true
  value of $\rho$;
\item GF0: the first coordinate $\hat{\alpha}$ of $\hat{B}$ (no 
  prior information);
\item FP: the fixed point estimator 
$\frac{\log(\hat{g}(\delta_*))}{\delta_*-1}$ (using the true value of $\delta$).
\end{itemize}
Table \ref{tab:estimalpha} shows the MSEs
obtained for different sets of parameters.
Not surprisingly, using the true value of $\delta$ or $\rho$ gives a
better estimate of $\alpha$;
the information on $\delta$ yields a better precision than the
information on $\rho$. Both estimators GFd and FP use the information
on $\delta$, but the first one is better. For low values of
$\alpha$, FP performs reasonably well, as does the $p_0$-method for
$\delta=0$. However for large values of $\alpha$, FP
is strongly biased. A value of $\rho$ smaller than $1$ implies a
heavier tail, hence larger and more frequent outliers. It worsens the
estimation of $\alpha$, whatever the estimator.

\begin{table}[!ht]
$$
\begin{array}{|c|cccc|}
\hline
(\alpha,\rho,\delta)&\mbox{GFd}&\mbox{GFr}&\mbox{GF0}&\mbox{FP}\\
\hline
(1,1,0.1)&0.040&0.087&0.136&0.041\\
(1,1,0.05)&0.039&0.076&0.144&0.041\\
(1,0.8,0.05)&0.042&0.105&0.166&0.044\\
(10,1,0.1)&0.272&0.506&1.000&2.677\\
(10,1,0.05)&0.894&1.143&1.983&11.84\\
(10,0.8,0.05)&1.000&1.393&2.088&13.87\\
\hline
\end{array}
$$
\caption{Mean squared errors on 4 estimators of $\alpha$ from 1000
  samples of size 1000 of the LDD$(\alpha,\rho,\delta)$ for different
  values of the parameters. The first estimate uses the true
  value of $\delta$, the second one the true value of $\rho$, the
  third one uses no prior
  information. The last one (fixed point) does not depend on $\rho$ and
  uses the true value of $\delta$.}
\label{tab:estimalpha}
\end{table}

Apart from simulation experiments, we have tried estimating
$\alpha,\rho,\delta$ on several samples of real data. The
results obtained on the 1102 data from 
Boe et al. (1994) are reported here.
These data
were ajusted on the LD$(\alpha,\rho)$ by
Zheng (2005)
using the maximum likelihood method: his estimates
of $\alpha$  and $\rho$ are $0.71$ and $0.84$ respectively. 
Table \ref{tab:boe} shows the estimates of $\alpha$ and $\rho$
obtained for values of $\delta$ ranging from $0$ to $0.3$. It also
gives the distances between the empirical distribution and the
estimated LDD, either in the sense of PGF's or in that
of cumulative distribution functions. The fit is
quite good, whatever the value of $\delta$. The value $\delta=0.06$ gives
the best fit in the sense of PGF's. Even though this
value is coherent with death probability 
estimates reported by 
Fontaine et al. (2008),
it cannot be
considered as reliable. Indeed, the 95\% confidence margin of error on
$\delta$, deduced from the asymptotic covariance matrix, is $\pm 0.13$. This
huge margin is coherent with what we have observed on simulated
samples with analogous parameters. 

\begin{table}[!ht]
$$
\begin{array}{|c|ccccccccccc|}
\hline
\delta&0.00& 0.03& 0.06& 0.09& 0.12& 0.15& 0.18& 0.21& 0.24&  0.27&
0.30\\
\hline
\hat{\alpha}(\delta)&0.71&0.72&0.73&0.75&0.77&0.79&0.81&0.83&0.87& 0.90& 0.95\\
\hat{\rho}(\delta)&0.82&0.83&0.84&0.84&0.85&0.86&0.87&0.88&0.90 &0.92& 0.94\\
DG&1.04&0.87&0.74&0.79&0.86&0.94&1.05&1.19&1.38& 1.66& 2.06\\
DF&6.48&6.25&6.29&6.53&6.79&7.08&7.41&7.79&8.23 &8.74 &9.35\\
\hline
\end{array}
$$
\caption{Estimates of $\alpha$ and $\rho$ for different values of
  $\delta$ on the data from 
Boe et al. (1994).
On row 4, DG is 
  the maximal distance between the EPGF function of the
sample and the PGF of the 
LDD$(\hat{\alpha}(\delta),\hat{\rho}(\delta),\delta)$, multiplied by
$10^3$. On row 5, DF is the distance between cumulative distribution
functions, also multiplied by $10^3$.}
\label{tab:boe}
\end{table}

The main conclusion of our
experimental study is that the death
probability $\delta$ cannot be reliably estimated on samples such that
the product $\alpha\times n$ is lower than $10^5$, which is far beyond current
fluctuation analysis experiments. We remark that available estimates
of $\delta$ have orders of magnitude of a few percents: see
Fontaine et al. (2008).
Table \ref{tab:distancesq0q1} for theoretical
distances as well as Table \ref{tab:boe} for actual data, permit to evaluate the
influence of $\delta$: it
turns out that the effect of a small $\delta$ 
 on the estimates of $\alpha$ and $\rho$ has the
same order of magnitude as $\delta$ itself. So neglecting the effect of cell
deaths if no reliable
estimate of their probability is available, seems legitimate.
\section{Conclusion}
A probabilistic model of fluctuation analysis, taking into account
cell deaths, has been proposed. A new family of distributions
LDD$(\alpha,\rho,\delta)$, modeling asymptotic number of mutants has been
derived. The three parameters are the expected number of mutations $\alpha$
(which is the parameter of interest in fluctuation analysis), the relative 
fitness of normal cells compared to mutants $\rho$, and the death probability
of mutant cells $\delta$. In the particular case $\delta=0$, the
classic Luria-Delbr\"uck distribution is recovered. The extension of
known mathematical results to the case $\delta>0$ is straightforward:
explicit simulation and computation algorithms for probabilities have been
described. The LDD$(\alpha,\rho,\delta)$ has the same type of asymptotic
behavior than the Luria-Delbr\"uck distributions: heavy tail with tail
exponent $\rho$. Thus, the occurrence of ``jackpots'' (large counts of
mutants) is a common feature. Modeling an observed sample of mutant
counts by a LDD$(\alpha,\rho,\delta)$ poses the problem of estimating
the three parameters simultaneously. If the death probability $\delta$
is known, then $\alpha$ and $\rho$ can be estimated using the
generating function method, exactly as in the
case $\delta=0$. The larger the expected number of mutations $\alpha$,
the more precise the estimates. However, for samples of size smaller
than $10^3$, all values of $\delta$ lead to a good fit, and the
different distributions so obtained can hardly be
distinguished. Choosing for $\delta$ the value giving the best fit
yields a consistent estimator with optimal mean squared error, but it
cannot be considered a reliable choice for small samples. Since the
death probabilities that have been reported in practice are small,
their influence on the estimates of $\alpha$ and $\rho$ can be
neglected as a first approximation, if no prior information on the
actual value of $\delta$ is available. A script containing the R
functions for the statistical treatment of the LDD distributions has
been made available online.
\section*{Acknowledgements}
The author thanks Anestis Antoniadis, Jo\"el Gaff\'e,
Agn\`es Hamon, Alain le Breton, and Dominique Schneider  for helpful
and pleasant discussions.


\end{document}